\documentstyle[prb,aps,floats,epsfig]{revtex} 
\newcommand{\bsim}{\mbox{\raisebox{-0.1cm}{$\;
\stackrel{\textstyle>}{\sim}\;$}}}
\newcommand{\lsim}{\mbox{\raisebox{-0.1cm}{$\;
\stackrel{\textstyle<}{\sim}\;$}}}
\begin{document}

\title{Band-filling effects on electron-phonon properties of
normal and superconducting state} 

\author{E. Cappelluti$^{1,2}$, S. Ciuchi$^3$, C. Grimaldi$^4$,
and L. Pietronero$^{2,5}$} 

\address{$^1$``Enrico Fermi'' Center,
v. Panisperna, 00184 Roma, Italy}

\address{$^2$Dipart. di Fisica, Universit\'{a} di Roma ``La Sapienza", 
Piazzale A.  Moro, 2, 00185 Roma,
and INFM UdR Roma1, Italy}

\address{$^3$ Dipart. di Fisica, Universit\'{a} dell'Aquila,
v. Vetoio, 67010 Coppito-L'Aquila, 
and INFM, UdR l'Aquila, Italy}

\address{$^4$ Institut de Production et Robotique, LPM,
\'Ecole Polytechnique F\'ed\'erale de Lausanne,
CH-1015 Lausanne, Switzerland}

\address{$^5$ Istituto di Acustica ``O.M. Corbino'',
CNR, Area di Ricerca Tor Vergata, Roma, Italy}

\maketitle 

\begin{abstract}
We address the effect of band filling on the effective electron mass $m^*$
and the superconducting critical temperature $T_c$ in a electron-phonon system.
We compare the vertex corrected theory with the non-crossing approximation
of the Holstein model within a local approximation. We identify two
regions of the electron density where $m^*$ and $T_c$ are 
enhanced or decreased by the inclusion of the vertex diagrams.
We show that the crossover between the enhancement at low density and 
the decrease towards half filling is almost independent of the
microscopic electron-phonon parameters. These different behaviors are
explained in terms of the net sign of the vertex diagrams which is positive
at low densities and negative close to half filling.
Predictions of the present theory for doped MgB$_2$,
which is argued to be in the low density regime, are discussed. 
\end{abstract}



\section{Introduction}
Recently, the discovery of superconductivity at $T_c\simeq 39$ K in magnesium
diboride MgB$_2$,\cite{akimitsu} has stimulated a strong interest on
the role of band-filling in phonon-driven
superconductivity when the charge carriers stem from low electron
or hole density in the conduction band. 
Classical examples of low density superconductors are doped
semiconductors like GeTe which display critical temperatures $T_c$
of the order of $1$ K or less.
Such small $T_c$ values are understood as basically due to low values
of the density of states at the Fermi level $N_0$
which goes to zero as $N_0 \propto \sqrt{E_{\rm F}}$,
where $E_{\rm F}$ is the Fermi energy.
In this case, the electron-phonon coupling, which scales with the
density of states, $\lambda=g^2 N_0$ where $g$ is the electron-phonon
matrix element, also vanishes as $\lambda\propto \sqrt{E_{\rm F}}$.
This situation changes drastically for systems having a
marked two-dimensional character of the conduction band at the Fermi level.
In fact, an ideal
two-dimensional system  has a finite value of $N_0$ even when
$E_{\rm F}\rightarrow 0$, leading in principle to a sizeable $\lambda$ 
also for very low charge carrier concentrations. 

Magnesium diboride belongs in principle to such class of materials.
In fact, the peculiarity of MgB$_2$ resides on its electronic 
band structure which is built up by three-dimensional $\pi$ bands 
and quasi two-dimensional $\sigma$ bands.\cite{belashchenko,an}
Both sets of bands have large bandwidth
of the order of $5$-$10$ eV, however the Fermi energy cuts the $\sigma$ band
quite near the band edge leading to $E_{\rm F}$ of only $0.4$-$0.6$ eV.
Since the Fermi level crosses the $\sigma$ bands around the $\Gamma$ point, the
low energy physics of the $\sigma$ bands is essentially captured by a simple
parabolic band structure corrected by a small interplane hopping
element.\cite{an}
The electron-phonon coupling has been estimated by several groups and
the common agreement is that the $\sigma$ bands are strongly coupled to lattice
vibrations of $E_{2g}$ symmetry and frequency $\omega_{\rm ph}$
of order $70$ meV, 
while the $\pi$ bands have a somewhat lower interaction.\cite{kong,bohnen,liu}
Hence, the main features of MgB$_2$ can be described by
a quasi-two dimensional 
parabolic band with low density of holes interacting with the lattice
vibrations.  

After the discovery of $T_c=39$ K in MgB$_2$,  the idea of varying the hole
density of the $\sigma$ bands to possibly enhance $T_c$ has been the object of
several works, both theoretical and experimental. A possible route is to
provide additional charge, electrons or holes, to the system by chemical
substitution of Mg or B atoms, although the rigid band scheme is somehow
questionable.  For instance,
by substituting Mg with Al it is possible to further decrease 
the Fermi energy of the pure MgB$_2$
\cite{MgAlB2_epxt1,MgAlB2_epxt2} and consequently to
explore the non adiabatic regime. It is interesting to see that
by increasing the Al content the critical temperature
decresase\cite{MgAlB2_epxt1,MgAlB2_epxt2} but not at the rate expected by first principles calculations carried
on with Eliashberg scheme\cite{profeta}.

An alternative way to chemical substitution is to
search for related compounds with similar electronic properties.
For example, insulating LiBC compounds have been
theoretically predicted to display high values of the superconducting critical
temperature $T_c$ for low hole-dopings corresponding to 
$E_{\rm F} < 0.5$ eV.\cite{pickett}

Such small values of $E_{\rm F}$ together with sizeable electron-phonon couplings
define a highly non trivial situation for which the classical Migdal-Eliashberg (ME)
theory becomes questionable. A major inadequacy of the classical theory
is the breakdown
of the adiabatic hypothesis occurring when the Fermi energy
becomes of the same order or smaller than the characteristic phonon
frequency scale $E_{\rm F} \lsim \omega_{\rm ph}$,\cite{migdal} or when
$\lambda$ becomes strong enough to 
give rise to polaronic effects.\cite{lang-firsov}
So far this issue has been
often associated with narrow band systems, as fullerenes,\cite{gunnarsson}
cuprates\cite{kulic} and manganites,\cite{manga}
where the Fermi energy is constrained
by the small electronic bandwidth.

As discussed above, however, the nonadiabatic
problem can be relevant also in large band materials with low carrier
density provided a non vanishing electron-phonon coupling $\lambda$.
The nonadiabatic degree $\omega_{\rm ph}/E_{\rm F}$ can be thus tuned
by varying the charge carrier density or alternatively the
chemical potential through the electronic band up to the top (or the bottom)
of the electronic states.

Aim of the present paper is to investigate
in details the onset of nonadiabatic effects in large band systems
as function of the charge density and their implications on the
superconducting and normal state properties.
We find two regions of the electronic filling in which nonadiabatic
contributions behave in qualitatively different manners: at sufficiently
low density they enhance the electron-phonon effects in normal and
superconducting state properties, while close to half-filling
nonadiabatic effects basically lead to a weakening.

\section{Nonadiabatic theory: non-crossing
and vertex correction approximations}
\label{general}

In this section we present a brief summary of the
theory of nonadiabatic superconductivity in term of the Green's function
formalism\cite{gpsprl} where the nonadiabatic
effects will be introduced at different stages of complexity.
An explicit analytical derivation of the superconducting equations
will be provided within the local approximation.

Let us consider a single band electron system 
interacting with phonons. In the normal state, the electron propagator is: 
\begin{equation}
\label{eq1}
G(k)^{-1}=i\omega_n-\epsilon_{{\bf k}}+\epsilon_0-\Sigma(k),
\end{equation}
where $\omega_n=(2n+1)\pi T$ is the fermionic Matsubara frequency,
$T$ is the temperature, $\epsilon_{{\bf k}}$ is the electron dispersion,
$\epsilon_0$ the chemical potential and $\Sigma(k)$
is the electronic self-energy. In Eq. (\ref{eq1}), the index
$k$ stands for the 4-momentum $({\bf k},\omega_n)$.
In the following, we assume that 
the Coulomb contribution to the normal state self-energy has already been
absorbed in the electron dispersion and the chemical potential. 
However, for the moment, we do not need
to specify the form of $\epsilon_{{\bf k}}$.
The electron self-energy 
$\Sigma(k)$ is determined by the electron-phonon interaction:
\begin{equation}
\label{eq2}
\Sigma(k)=\sum_{k'}V_{\rm N}(k,k')G(k'),
\end{equation}
where $\sum_{k'}=-(T/N)\sum_m\sum_{{\bf k}'}$. Up to the first order
of the irreducible electron-phonon interaction kernel of the self-energy
$V_{\rm N}(k,k')$
is given by:\cite{psg,gps}
\begin{equation}
\label{eq3}
V_{\rm N}(k,k')=V(k,k')+V(k,k')\sum_{k''}V(k,k'')G(k'')G(k''-k+k').
\end{equation}
In the above expression, the lowest order electron-phonon interaction
$V(k,k')$ is given by $V(k,k')=
g({\bf k},{\bf k}')^2 D(k-k')$
where $g({\bf k},{\bf k}')$ is the electron-phonon matrix element and $D(k-k')$ 
is the phonon propagator. In the following, we assume a phenomenological model for
$D(k-k')$, in the same spirit therefore of the Migdal-Eliashberg theory in which
the phononic degrees of freedom are taken from experiments while the electron
Green's function is calculated self-consistently.
Hence, we shall not consider the effects of the phonon self-energy. These are
however negligible at sufficiently low electron densities, that is the 
region of main interest to us.

If the second term in Eq. (\ref{eq3}) is neglected, the
so-called non-crossing approximation (NCA) is recovered,
which, in the limit of infinite bandwidth, is equivalent to the Migdal-Eliashberg
approximation.\cite{migdal,elia} 
When the full expression of Eq. (\ref{eq3}) is considered, 
the system is described within the vertex-corrected approximation (VCA),
equivalent
to the theory of nonadiabatic electron-phonon interaction discussed in 
Refs. \onlinecite{gpsprl,psg,gps}.
For later convenience, and to maintain a closer connection with
the notation used in Ref. \onlinecite{gpsprl,psg,gps},
we define the vertex function
\begin{equation}
\label{ver1}
P(k,k')=\sum_{k''}V(k,k'')G(k'')G(k''-k+k'),
\end{equation}
so that Eq. (\ref{eq3}) is simply 
\begin{equation}
\label{v2}
V_{\rm N}(k,k')=V(k,k')[1+P(k,k')].
\end{equation}

The extensions of NCA and VCA to describe the superconducting transition are 
readily obtained by either introducing the Nambu formalism or by computing the pairing
susceptibility. We adopt here the latter formalism so that the temperature $T_c$ at which
the pairing susceptibility diverges is given by the solution of the following secular equation:
\begin{equation}
\label{gap1}
\phi(k)=\sum_{k'}V_{\Delta}(k,k')G(k')G(-k')\phi(k'),
\end{equation}
where $\phi(k)$ is the off-diagonal self-energy and the 
kernel $V_{\Delta}$ is:
\begin{eqnarray}
\label{gap2}
V_{\Delta}(k,k')=&&V(k,k')+V(k,k')\sum_{k''}V(k,k'')G(k'')G(k''-k+k') \nonumber \\
&&+V(k,k')\sum_{k''}V(k'',k)G(-k'')G(-k''+k-k')
+\sum_{k''}V(k,k'')V(k'',k')G(k'')G(k''-k-k').
\end{eqnarray}
The NCA equations are obtained by neglecting all the
nonadiabatic corrections so that $V_{\Delta}(k,k')$ is simply $V(k,k')$. 
In this limiting case the electron-phonon interaction enters in the same
way in the interaction kernels of both the self-energy and the
superconducting pairing: $V_{\Delta}(k,k') = V_{\rm N}(k,k') = V(k,k')$.
This is however not true in the VCA framework where in principle
$V_{\Delta}(k,k') \neq V_{\rm N}(k,k')$.
In particular
the nonadiabatic terms add considerable structure to the pairing kernel but
symmetry considerations permit to simplify the structure of $V_{\Delta}(k,k')$.
In fact, if the system has inversion symmetry,  $V(k,k')=V(-k,-k')$ 
and $G(-k)=G(k)^*$
where $G(k)^*$ is the complex conjugate of $G(k)$. By using these relations 
it is easy to see that the third term on the right hand side of Eq. (\ref{gap2})
is the complex conjugate of the second term. Their sum is therefore twice
the real part of, for example, the second term.
Consider now the last term of Eq. (\ref{gap2}). If we replace the summed index
$k''$ with $-k''+k-k'$ then the resulting expression is equal to the
complex conjugate of the original term. Then also the last term of the
gap equation is real. These considerations permit to rewrite Eq. (\ref{gap2}) in the
more condensed form:
\begin{equation}
\label{gap3}
V_{\Delta}(k,k')=V(k,k')[1+2P_{\rm R}(k,k')]+C(k,k'),
\end{equation}
where $P_{\rm R}(k,k')$ is the real part of the vertex function $P(k,k')$ defined
in Eq.(\ref{ver1}), and
\begin{equation}
\label{cross1}
C(k,k')=\sum_{k''}V(k,k'')V(k'',k')G(k'')G(k''-k-k'),
\end{equation}
is the so-called cross correction, which is a real function of $k$ and $k'$.
Finally, reminding that $G(k')G(-k')=|G(k')|^2$, we can see that
the kernel of the linearized gap equation in Eq. (\ref{gap1})
is also real and without loss of generality
we can choose $\phi(k)$ as a real function of $k$.

In this article, we study both NCA and VCA as a function of band filling
for a model density of states (DOS) which is constant through the entire electron
bandwidth $D$. 
When the chemical potential $\epsilon_0$ is close to the top (bottom)
of the band, this model effectively mimics a hole (electron) doped two-dimensional
parabolic band, provided $\omega_{\rm ph}$ is smaller than the entire bandwidth.
This modelization is thus a good starting point to eventually
discuss filling effects in MgB$_2$.
Within the NCA framework, Marsiglio has already considered this model 
in his study of the superconducting instability.\cite{marsi}
What he found was that  the critical temperature $T_c$ decreases as the chemical potential 
$\epsilon_0$ approaches the bottom or the top of the band. Naively,
this reduction is expected since, close to the band edges, 
the average of the DOS over an energy window of order $\omega_{\rm ph}$  
gets reduced. Quantitatively, however, the reduction of $T_c$ is affected also by
non-rigid band effects induced by the electron-phonon coupling.

Within the theory of nonadiabatic superconductivity, this situation is drastically modified by
the non-trivial band-filling dependence of the nonadiabatic corrections.
To illustrate this point, let us consider the vertex function $P(k,k')$ defined
in Eq. (\ref{ver1}). In previous studies we have shown that,
at half filling, the
vertex correction has a marked dependence on the momentum 
transfer ${\bf q}={\bf k}-{\bf k}'$, being positive
for $v_{\rm F}|{\bf q}| < \omega_{\rm ph}$ and negative for $v_{\rm F}|{\bf q}| > \omega_{\rm ph}$, where
$v_{\rm F}$ is the Fermi velocity.\cite{gpsprl,psg,gps}
If there is no preference to forward scatterings,
the average over $|{\bf q}|$ of the vertex function is generally small and negative, 
reducing therefore the electron-phonon pairing and consequently
$T_c$ as compared to NCA.\cite{freericks,paci}
As the chemical potential is moved towards the top or bottom of the band, the 
${\bf q}$-dependence is more and more weakened and,
when the distance between the band edge and $\epsilon_0$ is sufficiently small,
the vertex function assumes 
an overall positive sign with negligible momentum dependence.\cite{pera}
In this regime
therefore the nonadiabatic corrections tend to enhance the pairing, and
the reduction of $T_c$ given by NCA is counter balanced
by the nonadiabatic corrections.
Close to the top or bottom of the band,
we expect
therefore a competition between band edge effects which depress $T_c$ and
nonadiabatic corrections which enhance $T_c$. Which of these two mechanisms
prevails over the other depends on the chemical potential, on the bandwidth 
and on the electron-phonon interaction strength. It is matter of this article to
provide quantitative results of this interesting problem.

\subsection{Local approximation}
\label{local}

To focus on the role of band-filling in enhancing or decreasing $T_c$, we employ some
approximations both in the description of the electron-phonon interaction and in the
way of dealing with the normal and off-diagonal self-energies.

First, we assume that phonon dynamics is well described by a single Einstein frequency 
$\omega_0$, so that the phonon propagator is
\begin{equation}
\label{eq4}
D(q)\equiv D(l)=-\frac{\omega_0^2}{\omega_l^2+\omega_0^2},
\end{equation}
where $\omega_l=2l\pi T$. In addition, we consider the electron-phonon matrix element
as independent of the momenta, {\it i.~e.},
$g({\bf k},{\bf k}') \equiv g$. This latter
assumption
disregards situations in which the electron-phonon interaction is strongly
momentum dependent such as in strongly correlated metals in which large momentum transfers
are suppressed by the rigidity of the correlated electronic
wave-function.\cite{kulic,grilli,zeyher}
Previous studies
have shown that in this situation the surviving small momentum scattering has deep
influences on the nonadiabatic corrections.\cite{gpsprl}
In order to focus on the role
of the band-filling effects,
however, we assume in the following
the electron-phonon matrix elements and the electronic Green's function
to be independent of the electronic momenta.
This assumption, which corresponds to the local approximation whose 
validity and limitations
will be discussed later, permits the comparison with an exact solution
in the limit of zero density. We note however that electronic
models interacting with purely local interactions
are widely studied as basis for the
recent developed dynamical mean-field theory which provides in some cases
an exact solution in infinite dimensions.\cite{georges,ciuk,si}
The local approximation can be viewed as an
average out of the momentum dependence over the whole Brillouin zone
of electron-phonon matrix elements
and of the electronic propagators.
Thus we formally replace the electron
propagators $G(k)$ by their local expressions
$G_{\rm loc}(n) = \sum_{\bf k} G({\bf k},n)$
which depend only upon the Matsubara frequencies $\omega_n$. The electron
self-energy becomes also momentum independent and can be rewritten as:
\begin{equation}
\label{eq5}
\Sigma(k)\rightarrow \Sigma(n)=i\omega_n-iW(n)+\chi(n),
\end{equation}
where $W(-n)=-W(n)$ and $\chi(-n)=\chi(n)$. With this notation, $G_{\rm loc}$
reduces to:
\begin{equation}
\label{eq6}
G_{\rm loc}(n)=\int d\epsilon
\frac{N(\epsilon)}{iW(n)-\epsilon+\epsilon_0-\chi(n)},
\end{equation}
where $N(\epsilon)$ is the bare electronic DOS and the integral over the energies
$\epsilon$ is performed over the whole bandwidth $D$.
We consider here a constant bare DOS system $N(\epsilon)=1/D$
[$-D/2 \le \epsilon \le D/2$]. The value of the chemical potential
$\epsilon_0=0$ corresponds thus to a half-filling case whereas, for zero electron-phonon
couplings, $\epsilon_0=-D/2$ and $\epsilon_0=D/2$ represent respectively
a completely empty and a completely filled band.
When $|\epsilon_0-D/2|\ll D$, this approximation simulates a quasi-two
dimensional parabolic band. By employing these assumptions, Eq. (\ref{eq6}) reduces to
\begin{equation}
\label{eq7}
G_{\rm loc}(n)=\frac{1}{D}\int_{-D/2}^{D/2}
d\epsilon\frac{1}{iW(n)-\epsilon+\epsilon_0-\chi(n)}
=-\frac{1}{D}[if(n)+g(n)],
\end{equation}
where we have defined
\begin{eqnarray}
\label{effe}
f(n)&=&\arctan\left(\frac{D/2-\epsilon_0+\chi(n)}{W(n)}\right)+
\arctan\left(\frac{D/2+\epsilon_0-\chi(n)}{W(n)}\right), \\
\label{gi}
g(n)&=&\frac{1}{2}\ln\left[\frac{W(n)^2+[D/2-\epsilon_0+\chi(n)]^2}
{W(n)^2+[D/2+\epsilon_0-\chi(n)]^2}\right].
\end{eqnarray}

By substituting Eq. (\ref{eq7}) into the definition of $V_{\rm N}$ of
Eq. (\ref{eq3}), the resulting interaction depends now only on Matsubara frequencies
and it reduces to:
\begin{equation}
\label{eq8}
V_{\rm N}(k,k')\rightarrow V_{\rm N}(n,m)=
g^2D(n-m)[1+\lambda P(n,m)],
\end{equation}
where $\lambda= g^2 N(0) = g^2/D$ is the electron-phonon coupling constant
and $P(n,m)$ is the vertex function:
\begin{equation}
\label{eq9}
P(n,m)=-DT\sum_l D(n-l) G_{\rm loc}(l)G_{\rm loc}(l-n+m)=
P_{\rm R}(n,m)+iP_{\rm I}(n,m),
\end{equation}
where
\begin{equation}
\label{PR}
P_{\rm R}(n,m)=-\frac{T}{D}\sum_l D(n-l)
[g(l)g(l-n+m)-f(l)f(l-n+m)]
\end{equation}
and
\begin{equation}
\label{PI}
P_{\rm I}(n,m)=-\frac{T}{D}\sum_l D(n-l)
[f(l)g(l-n+m)+f(l-n+m)g(l)]
\end{equation}
are the real and imaginary parts of the vertex function. Note that,
since $g(-n)=g(n)$ and $f(-n)=-f(n)$, Eqs. (\ref{PR}),(\ref{PI})
imply $P_{\rm R}(-n,-m)=P_{\rm R}(n,m)$ and $P_{\rm I}(-n,-m)=-P_{\rm I}(n,m)$.
Now, by applying
the local approximation to  Eq. (\ref{eq2}) and Eq. (\ref{eq5}) we can write in
a closed form the integral equations for $W(n)$ and $\chi(n)$:
\begin{eqnarray}
\label{W}
W(n)&=&\omega_n-\lambda T\sum_m D(n-m)\{[1+\lambda P_{\rm R}(n,m)]f(m)
+\lambda P_{\rm I}(n,m)g(m)\}, \\
\label{chi}
\chi(n)&=&\lambda T\sum_m D(n-m)\{[1+\lambda P_{\rm R}(n,m)] g(m)-
\lambda P_{\rm I}(n,m) f(m)\}.
\end{eqnarray}
The above equations are completed by the relation connecting the
electron occupation number $n$ with the chemical potential $\epsilon_0$:
\begin{equation}
\label{enne}
n=1-\frac{2T}{D}\sum_n g(n),
\end{equation}
where the factor $2$ stems from the
spin degeneracy. It is important to notice that, form Eq.(\ref{enne}), 
half-filling ($n=1$) is achieved
by $g(n)=0$, which implies that the imaginary part of the vertex function 
$P_{\rm I}$, Eq. (\ref{PI}), is zero. This is consistent with $\chi(n)=0$ and $\epsilon_0=0$
[see Eqs. (\ref{gi}),(\ref{chi})].

Finally we consider the local approximation in the superconductive instability
equations, Eqs. (\ref{gap1}),(\ref{gap2}).
By following the previous steps, and by
using Eqs.(\ref{eq7})-(\ref{gi}), it is straightforward to deduce that the gap
equation becomes momentum independent and that it reduces to: 
\begin{equation}
\label{gap4}
\phi(n)=-T\sum_m V_{\Delta}(n,m) f(m)\frac{\phi(m)}{W(m)},
\end{equation}
where the pairing interaction in the local approximation reads
\begin{equation}
\label{gap5}
V_{\Delta}(n,m)=\lambda[1+2\lambda P_{\rm R}(n,m)]D(n-m)+\lambda^2 C(n,m).
\end{equation}
In the above expression, $P_{\rm R}$ is the real part of the
vertex function given in Eq. (\ref{PR}), while $C(n,m)$ is the nonadiabatic
cross correction resulting from the local approximation of the last
term of Eq. (\ref{gap2}):
\begin{eqnarray}
\label{cross2}
C(n,m)&=&-TD\sum_l D(n-l)D(l-m)G_{\rm loc}(l)G_{\rm loc}(l-n-m) \nonumber \\
&=&-\frac{T}{D}\sum_l D(n-l)D(l-m)[g(l)g(l-n-m)-f(l)f(l-n-m)].
\end{eqnarray}

When Eqs.(\ref{effe},\ref{gi},{\ref{PR},\ref{PI},\ref{cross2}) are 
plugged into Eqs (\ref{W}-\ref{gap4}), we obtain a closed set of 
equations which can be numerically solved for both NCA and VCA to obtain
the normal state self-energy and the superconducting transition.
The numerical solution is achieved by using usual procedures:
first the normal state quantities $W(n)$, $\chi(n)$ are obtained by
iteration
by constraining the electron density $n$ to have a given value;
in a second step the gap equation, Eq. (\ref{gap4}),
is rewritten an an eigenvalue equation and $T_c$ is obtained 
when the highest eigenvalue becomes equal to the unity.

Let us stress that this theory could provide a route to analyze the data from
experiments via a theory which is valid even in a very low filling regime
where  Eliashberg and Mc Millan formula for $T_c$ are no longer valid. In fact
given an electron-phonon coupling  $g({\bf k},{\bf k}') $, as well as a
calculated phonon spectrum, it is possible to evaluate an effective local
interaction $V_{\Delta}$ by averaging over the {\it entire} Brillouin zone the
product $V(k,k')=g({\bf k},{\bf k}')^2 D(k-k')$ where $D(k-k')$ is the
corresponding phonon propagator. Notice that in this way filling effects will
enter {\it both} in filling dependence of el-ph coupling and of the
renormalized phonon frequencies. In this paper we conversely focus on the
explicit band-filling dependence of the Fermi energy which are the driving
effects at low density. To this aim we therefore choose a simple lorentzian
form for $V(k,k')$.

\section{Normal state properties}
\label{normal}

Our main aim is to investigate how the electron-phonon properties
(effective mass, superconducting pairing, etc\ldots)
depend, in both NCA and VCA, on varying the chemical potential as function of
the electronic filling in comparison to the phonon energy
scale $\omega_0$ and to the large electron energy scale provided
by the total bandwidth $D$.
In this section we focus on normal state properties while
the superconducting pairing will be discussed in Section \ref{critical}.
In relation to both these quantities
we shall distinguish two regimes:
one at low electron filling $n$ where vertex diagrams enhance the
effective electron-phonon coupling associated to the
renormalization of the electron mass and the particle-particle superconducting
pairing; one at large $n$ where
the renormalization due to the vertex processes leads to an effective
reduction of the coupling. We shall see that, while the magnitude
of these electron-phonon effects depends on the on the strength of
$\lambda$ and on the size of the adiabatic parameter $\omega_0/D$,
the typical filling where the crossover between these two regimes
occurs is only weakly dependent on the electron-phonon properties,
suggesting that it is mainly determined by purely electronic
reasons. The relation of these results with the sign and magnitude
of the vertex function will be discussed.

Concerning the normal state
we present results for two physical quantities: the electron density
as a function of the chemical potential and the effective mass ratio.
The first quantity is obtained through  Eq. (\ref{enne}) while the
effective mass ratio is evaluated as low energy limit
of the renormalized electronic frequencies [Eq. (\ref{W})]:
\begin{equation}
\frac{m^*}{m} = \left.\frac{W(n)}{\omega_n}\right|_{n=0}.
\label{eqmass}
\end{equation}
This expression provides a good estimation to the band mass as far as the
temperature is sufficiently low.
We have iteratively solved equations (\ref{W}),(\ref{chi})
at several temperatures
until we have approached the asymptotic low temperature mass ratio. 
In this case we found
that also the electron density is converged to its low temperature limit.
Calculations have been carried out at several values of
the chemical potential for very low
($\omega_0/D=0.01$) and moderately large
($\omega_0/D=0.1$) values of the phonon
frequency and for two values of the coupling constant representative of weak 
($\lambda=0.5$) and strong ($\lambda=1.0$) coupling.

\begin{figure}[t]
\protect
\centerline{\epsfig{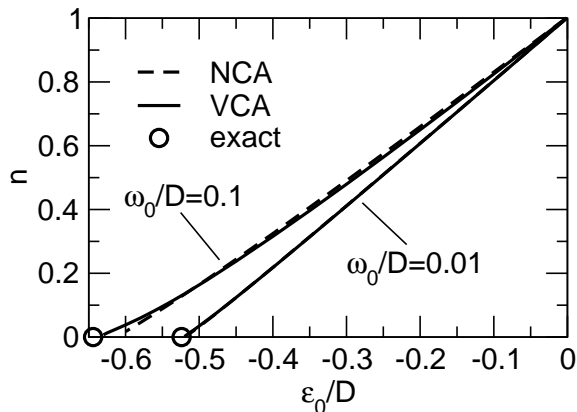}}
\caption{Electron density as a function of 
the chemical potential $\epsilon_0$ evaluated according NCA and the
VCA approximation for finite ($\omega_0/D=0.1$)
and almost zero ($\omega_0/D=0.01$) adiabatic parameter.
Empty circles represent the exact result of the zero density limit
evaluated by means of the continued fraction within the local theory.
Electron-phonon coupling is set $\lambda=1$.}
\label{fig_dens}
\end{figure}

In figure \ref{fig_dens} we report the behaviour of the electron density
as a function of the chemical potential evaluated using both NCA and VCA.
No significant difference between the two approximations is observed.
Note that, for small values of $n$, $\epsilon_0$ falls below the lower edge 
$\epsilon_0=-D/2$ of the bare conduction band.
This is especially evident when $\omega_0/D=0.1$
for which $\epsilon_0<-D/2$ already for
$n < 0.15$. 
In the zero density limit 
the chemical potential approaches
the bottom band edge
which corresponds to 
the ground state energy for a single electron
in interaction with phonons.
This energy
is lowered with respect to the bare value ($\epsilon_0/D=-0.5$) 
by the electron-phonon interaction.
Note that in the $n=0$ limit of this local theory
the electron and phonon properties
can be evaluated in a controlled way as well as in NCA and VCA by using
the continued fraction technique
(NCA and VCA correspond to different specific
approximations of the continued fraction).\cite{ciuk}
This is quite similar to the $d=\infty$ case where the local
ansatz is enforced by the infinite dimension limit.\cite{georges}
The zero density exact result of the continued
fraction is also reported in the figure (circles).
Note that, at least for these value of $\omega_0/D$ and $\lambda$
the vertex corrected theory provides already a rather good estimate of
the exact results at zero density.\cite{MaxPiovraCiuk}
We can expect thus a polaron crossover around some critical
value of the electron-phonon coupling $\lambda_c$ which depends
on the DOS shape. This feature however can not be reproduced
in NCA and VCA since it implies resummation of higher (infinite)
order vertex diagrams.\cite{ciuk}
The reliability of our approach is thus limited
to the metallic regime of the electron-phonon system here considered.
The value $\lambda_c \simeq 1.544$, corresponding to the polaron transition
in the adiabatic one particle electron-phonon system with rectangular
DOS, represents thus an evaluation of upper limit of validity of our analysis
near the adiabatic limit.
Note however 
that for $\lambda < \lambda_c$
there is no
reduction of the zero density ground state energy due to 
the electron-phonon
interaction
in the strictly adiabatic limit i.e.  when
$\omega_0/D\rightarrow 0$ \cite{Kabanov-Mashtakov}. 
This result, which includes all order vertex corrections,
holds true even when
the local approximation is relaxed.
Therefore we expect that frozen phonon approximation cannot reproduce 
this low density behaviour at least for couplings outside the polaronic regime.

After having discussed the $n$-$\epsilon_0$ phase diagram and
the range of validity of our analysis, we address now the filling
dependence of the effective mass ratio $m^* / m$.
\begin{figure}[t]
\protect
\centerline{\epsfig{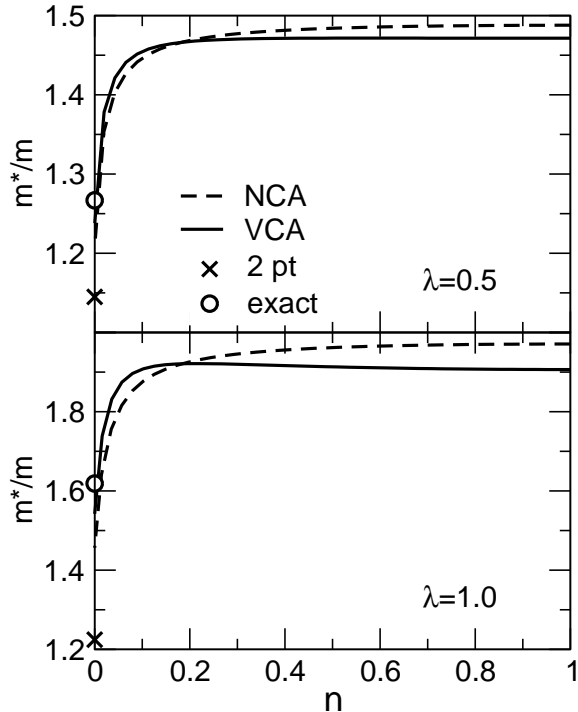}}
\caption{Effective
mass ratio for $\omega_0/D=0.01$ according the
non-crossing (NCA) and the vertex corrected (VCA) approximations.
In the zero electron density limit data are also shown
for exact calculations by means of continued fraction and
for the second order perturbation theory (2 pt).}
\label{fig_mw001}
\end{figure}
\begin{figure}
\protect
\centerline{\epsfig{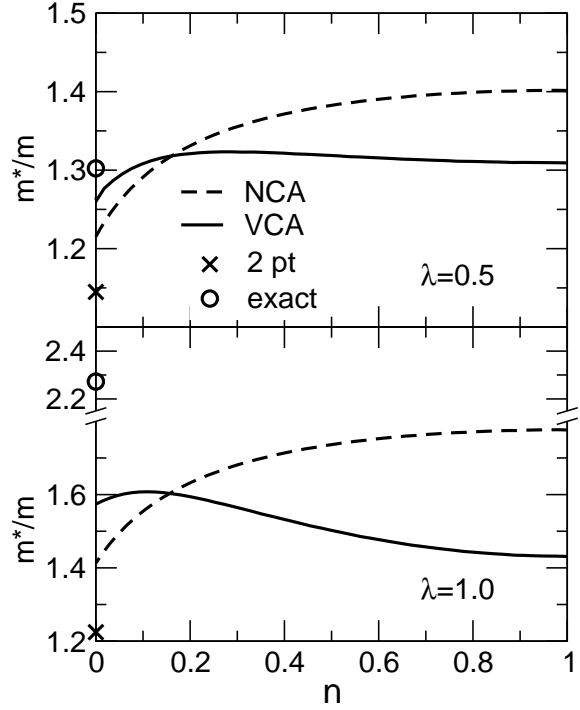}}
\caption{Effective mass ratio for $\omega_0/D=0.1$.
Notations are the same of Fig. \ref{fig_mw001}.}
\label{fig_mw01}
\end{figure}
In figures \ref{fig_mw001} and \ref{fig_mw01}
the results of the
calculations for $\omega_0/D=0.01$ and $\omega_0/D=0.1$
are respectively reported.
Because of band edge effects the effective electron mass $m^*$ is
reduced in NCA as the filling approaches $n \rightarrow 0$.
Increasing the phonon frequency leads to a broadening of this trend.
The introduction of vertex diagrams can affect this scenario.
In particular, as shown in Fig. \ref{fig_mw01} (lower panel),
for large enough electron-phonon coupling and adiabatic parameter
$\omega_0/D$ the effective electron mass $m^*$ in VCA can be {\em larger}
at the band bottom than at half-filling suggesting an increasing
of the effective electron-phonon coupling in this regime as due
to the vertex diagrams.

The comparison between NCA and VCA can provide an estimate of the
vertex effects alone.
In particular for both low and intermediate phonon frequencies
we can distinguish a low doping
regime in which the inclusion of 
the vertex diagrams increases the
effective mass \cite{MaxPiovraCiuk} and a high doping regime
in which effective mass is
reduced by the inclusion of the vertex corrections.\cite{gcp}

\begin{figure}
\protect
\centerline{\epsfig{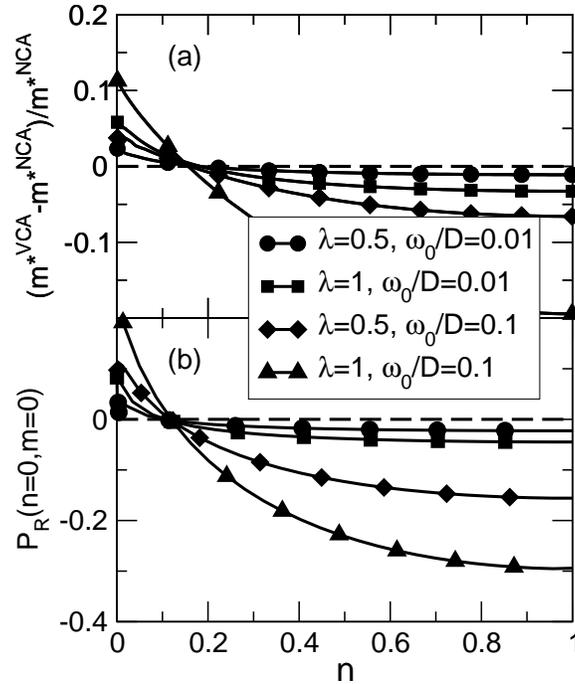}}
\caption{Panel (a): relative difference between effective
electron masses
evaluated according VCA and NCA as function of the electron filling.
Panel (b): sign of the real part of vertex function
$P_{\rm R}(n,m)$ [Eq. (\ref{PR})] evaluated at zero frequencies
$n=0$, $m=0$.}
\label{fig_diff_m}
\end{figure}

In order to single out the effects of the nonadiabatic diagrams
as function of the electron filling, we plot
in Fig. \ref{fig_diff_m}(a) the relative differences for VCA and NCA
results concerning the renormalization of the electronic mass $m^*$
[$(m^{\rm* VCA}-m^{\rm* NCA})/m^{\rm* NCA}$].
This plot shows clearly the
two regimes previously mentioned: a low electron density regime
where the introduction
of nonadiabatic vertex diagrams (VCA) leads to an enhancement of the
electron-phonon coupling with respect to NCA
for what concerns single particle properties ($m^*/m$);
and a large electron density regime
where the opposite holds true.
An interesting results is that the crossover filling
$n_c \sim 0.15-0.2$ where
nonadiabatic vertex effects change from positive
[$(m^{\rm* VCA}-m^{\rm* NCA})/m^{\rm* NCA} > 0$] to
negative
[$(m^{\rm* VCA}-m^{\rm* NCA})/m^{\rm* NCA} < 0$]
is almost independent of the bare electron-phonon coupling
$\lambda$ and of the adiabatic ratio
$\omega_0/D$.
We would like to stress again that for these
values of $n$ the chemical potential $\epsilon_0$, in particular
for $\omega_0 = 0.01 D$, can be quite far from the bottom of the band
with respect to the phonon energy, so that the crossover is not simply
related to band edge effects as described by a rigid band picture.

In order to better clarify the origin of this crossover from
a VCA $m^*$-enhancement region ($n \lsim 0.15-0.2$) to
a VCA $m^*$-reduction one ($n \bsim 0.15-0.2$), we plot in
Fig. \ref{fig_diff_m}(b) the size of the vertex function
$P_{\rm R}(n,m)$ [Eq. (\ref{PR})] evaluated in the static limit
$n=0,m=0$. Previous studies showed that the static limit
can be indeed representative of the effects of the vertex function
on static quantities, like $m^*$. The comparison of panels (a) and
(b) of Fig. \ref{fig_diff_m} points out striking similarities
between the filling behaviour of the two quantities.
It is thus easy to identify the sign of $P_{\rm R}$ as the direct origin
of the enhancement or suppression of the VCA electronic mass $m^*$
with respect to NCA. Once more we would like to stress that
while the magnitude of the discrepancy between NCA and VCA (panel a)
or $P_{\rm R}$ (panel b)
depend on the adiabatic parameter $\omega_0/D$ and on the
electron-phonon strength $\lambda$, the electron filling
$n_c \sim 0.15-0.2$ where vertex diagrams are almost perfectly balanced
is only weakly dependent on
electron-phonon properties.

An other interesting issue to be investigated is
the dependence of the electron-phonon renormalization on the
degree of the diagrammatic approximations employed.
Such a dependence
can be traced out in a controlled way in the zero electron density limit
where different theories
correspond to different degrees of diagrammatic approximations.
\cite{ciuk,MaxPiovraCiuk}
We compare thus the NCA and VCA with: $i$) the second
order perturbation theory (2 pt) where only the lowest order diagram with
bare propagators is taken into account,
namely in Eqs. (\ref{W})-(\ref{chi}) the r.h.s. is evaluated with
$W_n = \omega_n$, $\chi=0$ and $P_{\rm R}=P_{\rm I}=0$;
$ii$) and with the exact result for the local theory
obtained by the infinite resummation of the continued fraction.

For all the adiabatic parameters $\omega_0/D$ and coupling $\lambda$
here considered (Figs. \ref{fig_mw001}, \ref{fig_mw01})
we observe a monotonous increase of the effective mass
enhancement $m^* / m$ from the second order perturbation theory,
to NCA, VCA, ending up with the exact result.
This suggests that
vertex diagrams in the zero density limit
are positive defined
(the well defined positive sign of the vertex function in the zero density
limit was analytically pointed out in Ref. \onlinecite{gpscat} )
and the
resummation of higher and higher vertex diagrams,
as taken into account by the continued fraction,  converges uniformly
for the values of $\lambda$ and $\omega_0/D$ here considered.
In this situation the vertex corrected theory
can be considered qualitatively representative of the infinite
resummation exact result.\cite{MaxPiovraCiuk}
This picture is expected to be less
and less accurate
as soon as one approaches the polaronic regime.
For instance we
note that in the almost adiabatic case ($\omega_0/D = 0.01$)
VCA provides already a good estimation of the zero density
exact result even
for large coupling $\lambda = 1.0$. This can be understood by reminding that
electron-phonon polaron correlations are frozen in the adiabatic limit
up to $\lambda \lsim \lambda_c$, while they diverge at
$\lambda = \lambda_c$\cite{ciuk,Kabanov-Mashtakov}
On the contrary for $\omega_0/D = 0.1$ the approaching of
the polaron crossover is reflected in different properties (here for
instance in the electron mass renormalization) even for
$\lambda < \lambda_c$. This means that the qualitative trend from 2 pt
to NCA, VCA and to the exact result at $n=0$
is still preserved, but it lacks under the quantitative side as far
as $\lambda$ get closer to $\lambda_c$.\cite{MaxPiovraCiuk}

Summarizing the results of this section, we have identified two distinct
regions of the electronic filling where the vertex diagrams are respectively
positive and negative, leading to a corresponding increasing or reduction
of $m^*$ with respect to NCA. We have also shown that the extension
of this filling region does not vanish in the adiabatic limit.
The positivity of the vertex diagrams
can be better understood in the zero density limit.
This is a quite general result and
we believe that the physics of this limit can be qualitatively
representative of the low doping region ($n \lsim 0.15-0.2$)
at least in the case where
no phonon renormalization effects are not
explicitly taken into account
(note however that the phonon screening
vanishes as the charge density goes to zero).
Once again we stress that at large filling $n \bsim 0.15-0.2$
the system behaves in
a quite different way, since for instance the inclusion of vertex
diagrams leads to a {\em reduction} of the effective mass $m^*$
instead of an enhancement of it.

\section{Critical temperature}
\label{critical}

\begin{figure}
\protect
\centerline{\epsfig{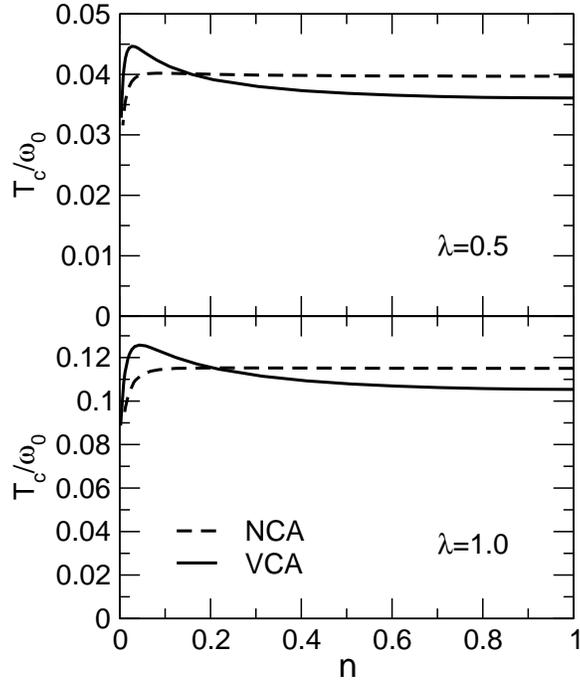}}
\caption{Critical temperature $T_c / \omega_0$
vs. the electron filling $n$
for $\omega_0/D=0.01$ in NCA and VCA.}
\label{fig_tcw001}
\end{figure}
\begin{figure}
\protect
\centerline{\epsfig{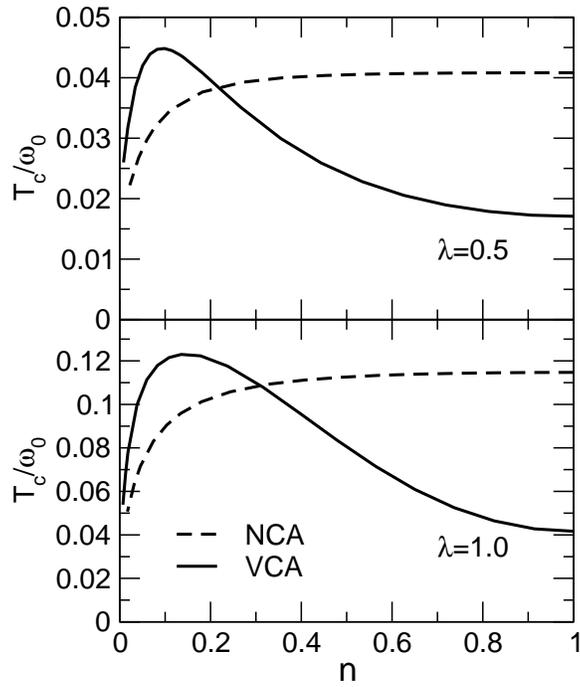}}
\caption{Same as Fig. \ref{fig_tcw001} for $\omega_0/D=0.1$.}
\label{fig_tcw01}
\end{figure}

The competition between band edge effects and nonadiabatic
contributions on the critical temperature shares quite similarities
with the case of the effective electron mass $m^*$.
In Figures \ref{fig_tcw001} and \ref{fig_tcw01} we report the results
for $T_c$ in both NCA and VCA frameworks, respectively
for $\omega_0/D=0.01$ and for $\omega_0/D=0.1$, as function of
the electron density $n$.

In analogy with the renormalized electron mass $m^*/m$,
the NCA results show that $T_c$ reduces as the electron occupation
number $n$ deviates from half-filling ($n=1$). 
This can be again naively understood in terms of reduced density
of states as the band edge is approached within
a phonon energy window. 
The resemblance with the $n$-dependence of $m^*/m$ is maintained also
when nonadiabatic corrections are included: at half filling $T_c$
is reduced with respect to NCA while, at sufficiently low $n$, $T_c$
is enhanced. This result is in agreement with previous studies on the
density effects on nonadiabatic corrections.\cite{freericks,pera}
At a first sight it could appear surprising that an enhancement of
the critical temperature is found in the presence of a corresponding
enhancement of the effective mass $m^*$ since strong renormalization
of $m^*$ is usually associated with a depletion of spectral weight.
However it was shown that even in the adiabatic theory in the
strong coupling limit  both  critical temperature   ($T_c \propto
\sqrt{\lambda}$) and effective mass ($m^*/m \propto \lambda$) increase.
\cite{Carbotte} 
In addition  it was pointed out
in Ref. \cite{ps} and later explicitly shown in
Refs. \cite{gpsprl,psg,gps}, that a different mechanism based on nonadiabatic
contributions leads to
a reduction of the spectral weight which is counterbalanced
by the additional opening of nonadiabatic interaction channels
in the Cooper paring even at intermediate coupling.

Although $m^*/m$ and $T_c$ behave in a qualitative similar manner,
interesting differences can be singled out by looking at
the relative change of $T_c$ due to the vertex diagram
inclusion: 
$(T_c^{\rm VCA}-T_c^{\rm NCA})/T_c^{\rm NCA}$ 
(Fig. \ref{fig_diff_tc}a).
\begin{figure}[t]
\protect
\centerline{\epsfig{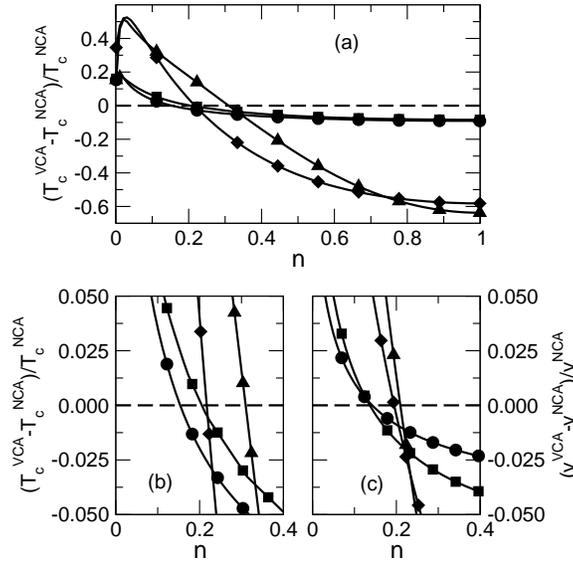}}
\caption{Panels (a): relative difference between $T_c$ evaluated
in VCA and in NCA $(T_c^{\rm VCA}-T_c^{\rm NCA})/T_c^{\rm NCA}$.
Panel (b):  relative difference between the maximum
eigenvector $v_{\rm max}$ of Eq. (\ref{gap5}) evaluated
in VCA and in NCA
$(v_{\rm max}^{\rm VCA}-v_{\rm max}^{\rm NCA})/v_{\rm max}^{\rm NCA}$.
Symbols as in legend of Fig. \ref{fig_diff_m}.}
\label{fig_diff_tc}
\end{figure}
We observe that the filling $n_c$ at which the inclusion of
vertex diagrams changes its character from positive to negative with respect
to the superconducting critical temperature is now quite sensitive of the
electron-phonon coupling $\lambda$ and of the adiabatic ratio
$\omega_0/D$. In particular $n_c$ increases as $\lambda$ and $\omega_0/D$
increase. This is in contrast with the case of the electron mass where
$n_c$ was almost independent of electron-phonon quantities.
There are two basilar reasons which account for this difference.
First, we remind that, while the electron mass renormalization is
in good approximation a {\em static} quantity (e.g. it is evaluated
at zero frequency), the calculation of $T_c$ involves the knowledge
of the electron-phonon kernel (and of the vertex function) in an
energy window $\sim \pm \omega_0$. The overall structure of the
vertex function can be thus remarkably dependent on
$\lambda$ and $\omega_0/D$ affecting the global balance for
which $T_c^{\rm VCA}=T_c^{\rm NCA}$.
An additional source of discrepancy comes from the fact that $T_c$ is
by definition a finite temperature quantity. The value of $T_c$ itself
is thus a $\lambda$-$\omega_0/D$ dependent quantities and finite
temperature effects can be drastically important in 
Fig. \ref{fig_diff_tc}a.

In order to single out these finite temperature effects we
evaluate the maximum eigenvalue $v_{\rm max}$ of the
superconducting electron-phonon kernel $V_\Delta (n,m)$
[Eq. (\ref{gap5})]
at low temperature in the normal state.
This provides thus an estimate of the superconducting pairing which is not
affected by finite temperature effects.
A zoom of the interesting region 
[$(v_{\rm max}^{\rm VCA}-v_{\rm max}^{\rm NCA})/v_{\rm max}^{\rm NCA}
\sim 0$] is shown in Fig. \ref{fig_diff_tc}c as well as
a corresponding zoom of 
$(T_c^{\rm VCA}-T_c^{\rm NCA})/T_c^{\rm NCA} \sim 0$.
As a first remark we note that, contrary to
Fig. \ref{fig_diff_tc}b, in Fig. \ref{fig_diff_tc}c
there is only a weak dependence of $n_c$
on the electron-phonon coupling $\lambda$. As matter of fact the
dependence on $\lambda$ in Fig. \ref{fig_diff_tc}b mainly stems
from finite temperature effects, with a large $\lambda = 1.0$
associated with higher temperatures ($T_c \simeq 0.1-0.12$)
in comparison of $T_c \simeq 0.03-0.04$ for small $\lambda=0.5$.
Once disregarded finite temperature effects, 
we still note a significant dependence on
the adiabatic ratio $\omega_0/D$ in Fig. \ref{fig_diff_tc}c which
does not appear for instance in Fig. \ref{fig_diff_m}.
As above discussed,
we trace back this residual dependence on $\omega_0/D$
to the important role of the frequency dependence
in the superconducting kernel.
Interesting, in the almost adiabatic regime $\omega_0/D = 0.01$,
the same value of $n_c \simeq 0.15$ as found for
the electronic mass and for $P_{\rm R}(n=0,m=0)$ is recovered,
signalizing the static nature of that value.
It is significant that both taking into account the retarded
nature of the superconducting pairing and finite temperature
effects yield an increase of the low doping region where
nonadiabatic vertex diagram enhance the electron-phonon coupling.
The value $n \lsim 0.15-0.2$ can be thus considered as
a conservative estimate of this regime.

In order to focus on the effects of nonadiabatic electron-phonon contributions,
we have not considered in this work the role of Coulomb repulsion between electrons.
In the simplest scheme, the electron-electron repulsion can be included by adding an
Hubbard $U$ term in Eq.(\ref{gap1}), as done in Ref.\onlinecite{marsi}. By treating $U$ as
a parameter independent of electron filling we have performed calculations of $T_c$
in both NCA and VCA. We have found  that the inclusion of $U$ leads to an overall
reduction of $T_c$ and to a magnification of the differences between NCA and VCA
reported in this section. However, no qualitatively new physics is induced by
$U$, although we expect an additional reduction of $T_c$ 
when the electron density is sufficiently low
to poorly screen the Coulomb interaction. However, the role of screening was not 
the aim of this paper, and its effects will be investigated in a future publication.

\section{Discussion and conclusions}
\label{concl}

In this paper we have addressed the role of band filling
in some normal and superconducting properties
of electron-phonon systems. 
In particular, within a local approximation,
we have identified two regions of the electron band filling
in which electron-phonon properties behave in
qualitatively different manners. At sufficiently low density, 
the inclusion of vertex diagrams beyond the non-crossing approximation 
leads to a strengthening of the electron-phonon effects reflected by
an enhancement of the effective electron mass $m^*$ and of the
superconducting critical temperature $T_c$. On the contrary, close
to half filling, the vertex corrected approximation results in
a net decrease of $m^*$ and $T_c$ with respect to the non-crossing
approximation. We have shown that these different behaviors are
governed by the net sign of the vertex diagrams which is positive in
the low density regime and negative close to half-filling.
The density value $n_c \simeq 0.15-0.2$
separating these two regimes depends only
weakly on the microscopic electron-phonon parameters which, on the other
hand, rule the size of the corresponding enhancement or decrease of
$m^*$ and $T_c$.

The results for the low electron density regime are compared with
the zero density limit for which the exact solution of this local
problem is known. In this limit we have shown that the inclusion
of diagrams with higher and higher order, from second order perturbation
theory to the non-crossing approximation and to the vertex corrected theory,
monotonically approaches the exact results. Such an
improvement as function of the degree of approximation suggests that
the positive sign of the vertex function leads to a
series with positive terms, which should steadily approach the exact
solution if a resummation of infinite order diagrams were
performed.\cite{MaxPiovraCiuk}

The enhancement of the
effective electronic mass $m^*$ and of the superconducting critical
temperature $T_c$
in the low electron density region
is expected to be relevant in twodimensional systems where
small charge density is accompanied by a finite density of states
at the bottom (or top) of the band. Thus, in principle, our results
could concern the  high-$T_c$ compound MgB$_2$  and
the hypothetical doped LiBC recently proposed to superconduct
with even higher $T_c$.\cite{pickett}
In particular in MgB$_2$ the hole carrier density
related to the $\sigma$ bands $n_h\simeq 0.08 < n_c$\cite{an} sets the system 
in a region where these effects are relevant,
although the large bandwidth of the $\sigma$ bands in MgB$_2$
compared with the relevant phonon frequency leads to $\omega_0/D\sim 0.01$,
corresponding to a quite weak effect on $m^*$ and $T_c$.
These effects can be however
further enhanced via Al substitution. Interestingly,
the experimentally observed
reduction of $T_c$ as function of Al doping \cite{MgAlB2_epxt1,MgAlB2_epxt2}
cannot be
accounted within a McMillan scheme \cite{profeta} even when
the Coulomb pseudopotential $\mu^*$ is introduced
to account for $T_c \simeq 39$ K in the undoped MgB$_2$ compound.

A quantitative enhancement of $T_c$ from vertex corrections in MgB$_2$ 
could be argued by considering two distinct effects not considered in 
the present paper namely: 

$a$) the intrinsic momentum dependence of the {\it bare} electron-phonon 
matrix 
elements;
$b$) the momentum dependence of the renormalized electron-phonon vertex. 

Let us discuss them separately. 

$a$) Several first principle calculations have shown that the electron-phonon 
coupling in MgB$_2$ has a strong momentum dependence and is particularly 
strong 
near the $\Gamma$ point at ${\bf q}=0$.\cite{kong,bohnen} In this 
situation, 
neglecting the momentum dependence of the electron-phonon matrix 
elements 
$g_{\bf k,k'} \sim g$ becomes a poor ansatz just as in the case of 
strongly 
correlated electron systems.\cite{kulic,grilli,zeyher} A proper 
inclusion of 
the predominance of forward scattering at small ${\bf q}$ has shown to 
lead to 
an intrinsic selection of the positive sign for the vertex diagrams and 
to 
favor a corresponding enhancement of $m^*$ and $T_c$.\cite{gpsprl} 

$b$) Even neglecting the momentum dependence of the {\it bare} 
electron-phonon 
matrix elements the renormalized vertex acquires a momentum dependency 
due to 
vertex corrections. The local approximation treat these effects  by 
averaging 
the intrinsic dependence of the vertex diagrams on the momentum transfer 
${\bf  q}={\bf k}-{\bf k'}$. In extreme cases the local assumption can 
ill-estimate 
the total phase space available for these vertex processes and hence the 
magnitude of the vertex diagrams. This is the most evident for instance 
in the 
low density adiabatic regime where the available transferred momenta are 
limited, in weak coupling regime,
by $\sim |{\bf q}| \le 2k_{\rm F}$, while the local approximation would 
integrated them over the whole Brillouin zone. In this limit for 
instance it is 
easy to show that the local approximation provide an underestimate for 
the magnitude of the vertex diagrams. 

Both effects could be separately considered to account properly of 
vertex 
corrections in MgB$_2$. Work in this direction is in progress.

This work was partially supported by the projects COFIN-2001 MIUR
and PRA-UMBRA INFM. We thank G. Profeta for useful discussions.

\end{document}